\documentclass{ieeeaccess}
\usepackage{cite}
\usepackage{amsmath,amssymb,amsfonts}
\usepackage{graphicx}
\usepackage{textcomp}

\usepackage{algorithm}
\usepackage{listings}
\usepackage{lipsum}
\usepackage{multirow}
\usepackage{lineno,hyperref}
\usepackage{algpseudocode}
\usepackage{booktabs}
\usepackage{mathtools}
\usepackage{upgreek}

\usepackage{soul}  

\usepackage{caption}
\DeclareCaptionFont{ieeeblue}{\color{accessblue}}
\DeclareCaptionLabelFormat{myformat}{\raggedright\figcapfont{\textbf{#1}\textbf{#2}}}
\def\BibTeX{{\rm B\kern-.05em{\sc i\kern-.025em b}\kern-.08em
    T\kern-.1667em\lower.7ex\hbox{E}\kern-.125emX}}
\begin{document}

\history{Received October 31, 2020, accepted November 24, 2020, date of publication November 27, 2020, date of current version December 11, 2020}
\doi{10.1109/ACCESS.2020.3041084}

\title{Detection of financial opportunities in micro-blogging data with a stacked classification system}
\author{\uppercase{Francisco de Arriba-P\'erez}\authorrefmark{1},\uppercase{Silvia Garc\'ia-M\'endez}\authorrefmark{1},\uppercase{Jos\'e A. Regueiro-Janeiro}\authorrefmark{1},\uppercase{ and Francisco J. Gonz\'alez-Casta\~no\authorrefmark{1}}}
\address[1]{Information Technology Group, atlanTTic, School of Telecommunications Engineering, University of Vigo, Campus, 36310 Vigo, Spain}
\tfootnote{This work was partially supported by Ministerio de Econom\'ia, Industria y Competitividad, Spain [TEC2016-76465-C2-2-R]; and Xunta de Galicia [GRC2018/053,ED341D-R2016/012].}

\markboth
{Francisco de Arriba-P\'erez \headeretal: Detection of financial opportunities in micro-blogging data with a stacked classification system}
{Francisco de Arriba-P\'erez \headeretal: Detection of financial opportunities in micro-blogging data with a stacked classification system}

\corresp{Corresponding author: Francisco J. Gonz\'alez-Casta\~no (e-mail: javier@gti.uvigo.es).}

\begin{abstract}
Micro-blogging sources such as the Twitter social network provide valuable real-time data for market prediction models. Investors' opinions in this network follow the fluctuations of the stock markets and often include educated speculations on market opportunities that may have impact on the actions of other investors. In view of this, we propose a novel system to detect positive predictions in tweets, a type of financial emotions 
which we term ``opportunities'' that are akin to ``anticipation'' in Plutchik’s theory. Specifically, we seek a high detection precision to present a financial operator a substantial amount of such tweets while differentiating them from the rest of financial emotions in our system. We achieve it with a three-layer stacked Machine Learning classification system with sophisticated features that result from applying Natural Language Processing techniques to extract valuable linguistic information. Experimental results on a dataset that has been manually annotated with financial emotion and ticker occurrence tags demonstrate that our system yields satisfactory and competitive performance in financial opportunity detection, with precision values up to $83$\%. This promising outcome endorses the usability of our system to support investors' decision making.
\end{abstract}


\titlepgskip=-15pt

\maketitle

\section{Introduction}

\subsection{Motivation}

Among the many factors that compel people to take decisions in stock markets, opinions on micro-blogging sites deserve consideration \cite{Li2018,Li2020,Bello-Orgaz2020}. It has been shown that fast paced information in social media is valuable for sales prediction \cite{Meire2017,Pai2018,Yuan2018} and has strong influence on micro-economic trends in stock markets \cite{Mai2018}. Investors’ decisions can be affected not only by media content \cite{Li2018} but also by public mood \cite{Sun2020}. Besides, the advent of powerful social trading platforms, such as eToro\footnote{Available at {\tt https://www.etoro.com/}, August 2020.} and {\sc xtb} online trading\footnote{Available at {\tt https://www.xtb.com/}, August 2020.}, has broadened the spectrum of input data. There already exist online tools to analyse these sources. Two examples are Thomson Reuters Eikon\footnote{Available at {\tt https://eikon.thomsonreuters.com/}, August 2020.}, which performs sentiment analysis ({\sc sa}) of social media content, and the Bloomberg platform\footnote{Available at {\tt https://www.bloomberg.com/}, August 2020.}, which provides social media indicators.

Furthermore, \cite{Enke2005} argued that, even though stock markets contain enough information to define their state, this information becomes obsolete before the general public can process it to take decisions. Therefore, instantaneous information in the Twitter micro-blogging network deserves attention \cite{Gerber2014,Reece2017,Zahra2020}, and, in particular, in finance \cite{Oliveira2017,Mai2018}. 

Stock market data mining conforms a significant body of research \cite{Nofer2015,Dimpfl2016,Zhong2017,Zhang2018,Hoseinzade2019} supporting market predictability. Some works have focused on Natural Language Processing ({\sc nlp}) techniques \cite{Sun2014,Fisher2016,Xing2018} and their application to emotion analysis ({\sc ea}) \cite{singh2016score,Razi2017}.

To the best of our knowledge, this work is the first proposal based on {\sc ea} techniques to detect what we call financial opportunities, that is, Twitter posts that speculate or reason that the value of particular actives will grow, rather than merely expressing positive opinions about those actives. Specifically, we present a three-layer stacked classifier with a first stage to discard neutral entries, a second stage to distinguish generic positive emotions from negative ones and a last stage to differentiate opportunity entries --our ultimate goal-- from any other positive emotions.

For this purpose, the system is enhanced with sophisticated linguistic information features such as $n$-gram sequences, emotion and polarity dictionaries, and frequency counters of hashtags, numerical information and percentages.

\subsection{Research goal and contributions}

We seek a Machine Learning system to detect financial opportunities in textual information extracted from tweets, where by opportunities we refer to positive user speculations and forecasts (e.g. ``the average yield of Google is 15.5\%'') instead of mere positive statements (e.g. ``I like Google''). They are akin to looking forward positively for upcoming events, as “anticipation” in Plutchik’s theory \cite{Plutchik2004}. 

This problem has not been previously considered in academic research and has remarkable practical applications for both stock market screening and decision making. Notwithstanding this applied perspective, our research has also a theoretical goal, an enhanced understanding of the linguistic dimension of micro-blogging comments to extract investment opportunities. Summing up, this work contributes to more effective Machine Learning classifiers for stock market data mining with the following:

\begin{itemize}
 
 \item A three-layer stacked Machine Learning classification system for detecting opportunities in micro-blogging comments. 
 \item An analysis of the most relevant features through {\sc nlp} techniques.
\end{itemize}

In addition, we employ a new dataset of investors' comments for stock market forecasting with 6,000 entries --similar in size to those in other state-of-the-art studies \cite{Chatzis2018,Al-Smadi2019,Simester2019,Tuke2020}--, which has been manually annotated with financial emotion and ticker occurrence tags (e.g. in ``the average yield of Google is 15.5\%'' the ticker tag is GOOGL) by experts in the field.

The rest of this article is organised as follows. Section \ref{sec:related_work} reviews related work in knowledge extraction, discussing {\sc ea} from general and financial perspectives. Section \ref{sec:system} describes the classification problem and our solution. Section \ref{sec:preliminaryTest} presents the experimental text corpus and the numerical tests that validate our approach. Finally, Section \ref{sec:conclusions} concludes the paper.

\section{Related work}
\label{sec:related_work}

Novice investors in stock markets give high consideration to the experience of financial experts. Thus, automatic knowledge extraction from the comments by experts in financial news, blogs and social media is an interesting research goal and a valuable asset for practical applications \cite{Rickett2016,He2017}.

Previous researchers have analysed online sources to characterise stock markets. Most of them have focused on news, as \cite{Alanyali2013}, who studied the correlation between financial market events and financial news content. In this regard, the work by \cite{Atkins2018} deserves a special mention. Using Machine Learning models of Latent Dirichlet Allocation, they concluded that the information extracted from news sources is better than price variations for predicting assets' volatility. Additionally, in \cite{Day2016} a Deep Learning ({\sc dl}) model for news categorisation was presented proving that financial sources of information have a significant effect on investments. There is also work on knowledge extraction from financial blogs. For example, the method by \cite{Wang2017} combined stock indexes with sentiment time series from micro-blogs. Besides, \cite{Rickett2016} applied multivariate regression to financial blogs to study how the markets react to the posts. Finally, the study of the information posted on social media platforms like Twitter has been a relevant research topic in recent years. Consider for example the analysis of consumer profiles by \cite{Ioanas2014} and the study of the evolution of stock prices and social media content by \cite{Sun2016} based on a latent space model \cite{Ming2014}. However, none of these works has considered user speculations about the evolution of the assets. Note that all of them have taken temporal information directly from the timestamps of financial forecasts or posts \cite{Rickett2016,Sun2016,Wang2017,Atkins2018}. In fact, we are not aware of any financial research supported by temporal analysis at discursive level such as ours\footnote{This aspect is relevant. Even though \cite{Gibbs1998} already noted the interest of the temporal perspective of the customers, i.e. the importance of their own understanding of time, few works have addressed temporal analysis at discursive level. Among early ones we must mention \cite{Forray2005}, which extracted time from punctuation, word choices and academic terminology or keywords in journal titles, by also considering common sense context information.}.

In this context, {\sc nlp} techniques \cite{liu2015sentiment} such as {\sc sa} \cite{Derakhshan2019} and {\sc ea} \cite{Staiano2014,Ge2020} have successfully extracted knowledge from comments of stock market experts. Specifically, in \cite{Staiano2014} the authors presented a novel approach for the automatic extraction of a high-coverage and high-precision lexicon annotated with emotion scores, while in \cite{Ge2020} they confirmed the impact of social media emotions on the stock market.

Often as an initial stage of {\sc ea}, {\sc sa} infers sentiment polarity from language features of user opinions \cite{Xu2019,Derakhshan2019}. Typically, three polarity levels (negative, neutral and positive) are considered \cite{Fang2018}, although some authors add two deeper positive and negative sentiment levels \cite{Balahur2015SentimentTweets, buvcar2016sentiment}. Among the wealth of research in Twitter mining, we can cite for example the work by \cite{Zimbra2016} on the analysis of consumer opinions about commercial brands, or by \cite{Smailovic2013PredictiveApplication} on financial {\sc sa}. The latter classified Twitter information about eight stock markets with a Support Vector Classifier ({\sc svc}). 

{\sc ea} infers feelings from linguistic information in user comments \cite{Rout2018,Chen2018}. It has become a central topic in the field of Affective Computing. Some typical emotion classes in the literature are love, joy, anger, sadness, fear and surprise, as defined by \cite{Parrott2001EmotionsReadings}. For example, the approach by \cite{neviarouskaya2007textual} detected affection in virtual communication environments. {\sc ea} has numerous applications in education \cite{Xu2018}, healthcare \cite{Asghar2017,Bong2017} and gaming \cite{ShamimHossain2015}, to cite some. It has also caught the attention of the industry as a profitable asset. In the particular case of finance, \cite{Sanchez-Rada2014} proposed a Linked Data approach for modelling sentiment and emotions from tweets about securities of the Spanish stock market. The system for financial forecasting by \cite{singh2016score} incorporated different sources of information into an integrated river model and combined {\sc gpoms}\footnote{Google Profile of Moods States ({\sc gpoms}). At the time this article was written it was no longer available.} with a Neural Network classifier, and \cite{Razi2017} applied a multi-layer perceptron to study the influence of investor emotions in investment decisions. At the end, even though {\sc ea} is still an interesting research topic \cite{Duxbury2020,Pengnate2020}, no previous analyses have taken into consideration finance-related emotions,
such as opportunity or anticipation, and have focused exclusively on general human emotions.

Training methodologies, on the other hand, can be divided into supervised (manually annotated), semi-supervised and unsupervised approaches. For example, in \cite{Fernandez-Gavilanes2018CreatingDescriptions} some authors of this paper proposed an unsupervised system for automatic generation of emotion lexica with polarity data. Hybrid solutions such as \cite{alvarez2015gti} (which applied classifiers as well as unsupervised approaches with polarity lexicons and syntactic structures) have produced high competitive results. Finally, stacking strategies combining different Machine Learning techniques into the same predictive model can improve performance \cite{Mehmood2018,Wang2019}. It has been our intent to exploit the benefits of stacking in ensemble learning techniques to further increase the performance of our system.

\section{System architecture}
\label{sec:system}

\begin{figure*}[!htbp]
 \centering
 \includegraphics[scale=0.25]{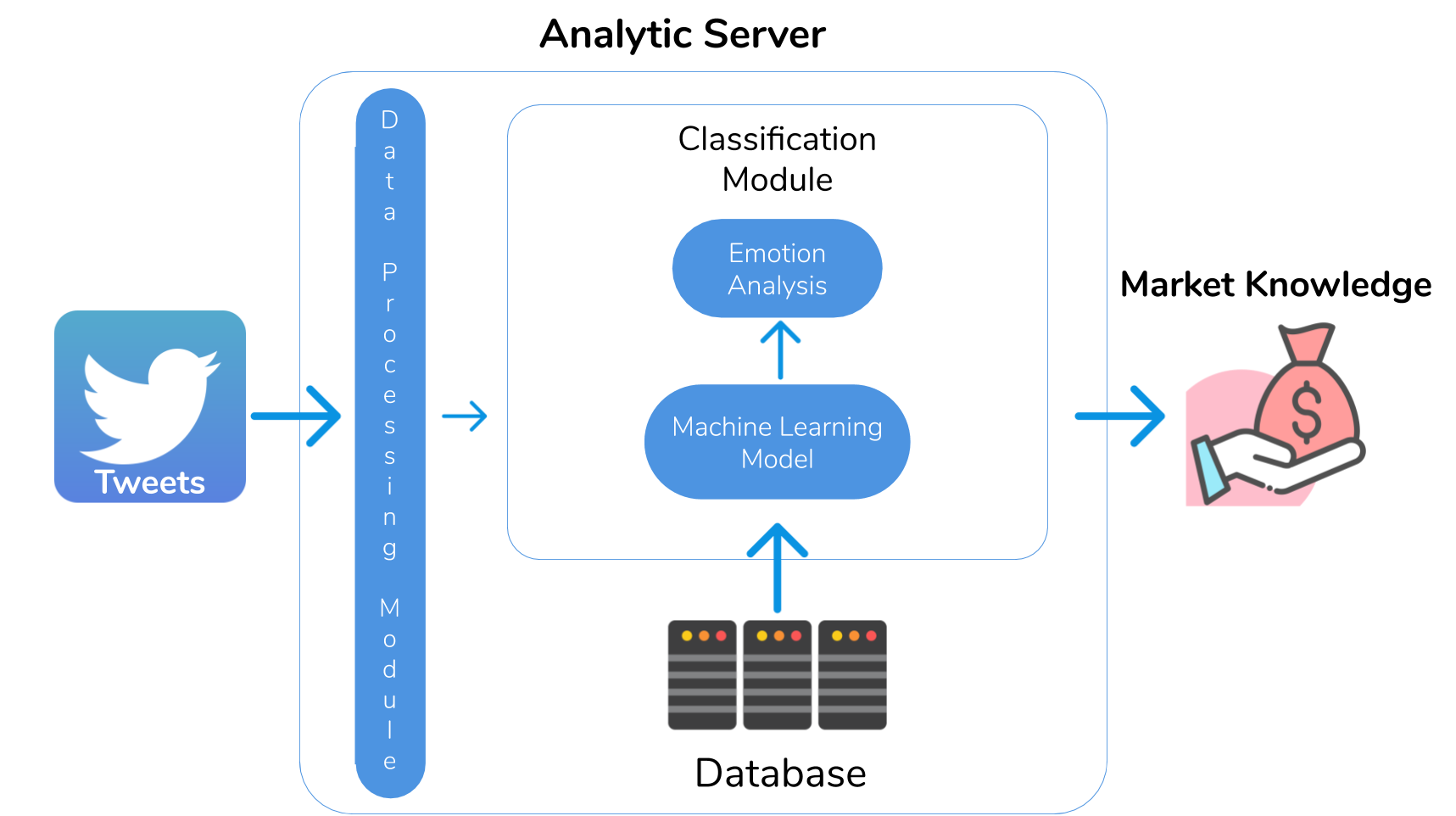}
 \caption{Our proposed framework.}
 \label{fig:system_architecture}
\end{figure*}

In this section we present our novel three-layer stacked system for detecting financial tweets on market opportunities. Specifically, the first stage detects neutral entries, the second stage distinguishes between general positive and negative emotions and the last stage separates opportunity entries from any other positive statements. Figure \ref{fig:system_architecture} illustrates its framework, which consists of two modules that will be described in detail in the next subsections, where financial tweets are taken from a training database. Table \ref{tab:entries_dataset} shows examples of input data and the corresponding emotion categories (including ``opportunity'', which we have already described). In the context of Plutchik's theory \cite{Plutchik2004}, ``positive statement'' corresponds to any positive emotion on present events such as ``expectation''; and ``negative awareness'' to any negative opinion about an asset, regardless of the sentiment expressed in the tweet, which is similar to ``disapproval''.

\begin{table*}[!htbp]
\centering
\caption{Examples of dataset entries with emotion tags.}
\begin{tabular}{lcc}
\hline
\multicolumn{1}{c}{\textbf{Tweet}} & \textbf{Emotion} \\ \hline

\begin{tabular}[c]{@{}l@{}}\textit{Parece que el \#IBEX35 va a hacer un buen cierre diario,}\\ \textit{aunque me gustaría que fuese por encima de los 8.600€.} \\`It seems that \#IBEX35 is going to make a good daily close, \\although I would like it to break 8,600€.'\end{tabular} & \begin{tabular}[c]{@{}c@{}}Opportunity\end{tabular} \\ 
& &\\

\begin{tabular}[c]{@{}l@{}}\textit{ \#FelizLunes ultimo día del año y el \#ibex35 subiendo de chiste :)} \\ `\#HappyMonday last day of the year and \#ibex35 is going up jokingly :)'\end{tabular} & \begin{tabular}[c]{@{}c@{}}Positive\\statement\end{tabular} \\
& &\\

\begin{tabular}[c]{@{}l@{}}\textit{\#DowJones cierra con pérdida de 0.36 por ciento} \\ `\#DowJones closes with a 0.36 percent loss' \end{tabular} & \begin{tabular}[c]{@{}c@{}}Negative\\awareness\end{tabular} \\ 
& &\\

\begin{tabular}[c]{@{}l@{}}\textit{Ya están los resultados de \$NFLX ???} \\ `Are \$NFLX results available ???'\end{tabular} & Neutral
\\ \hline
\end{tabular}

\label{tab:entries_dataset}
\end{table*}

\subsection{Data processing module}

This module discards useless data. Specifically, we apply a series of text transformations to ensure the quality of the data that enter the classification module. Table \ref{tab:beforeafterpreproOriginal} presents some examples before and after applying these transformations.

\begin{itemize}
 \item Filtering. It selects finance-related tweets while discarding spam and related tweets. Basically, tweets containing asset identifiers such as stock markets' tickers, {\sc url}s and quantitative information. We also remove advertising entries with spam-related words and expressions such as {\em sorteo} `draw', {\em gratis} `free' and {\em ?`quieres ganar dinero?} `do you want to earn money?'. Moreover, we discard words in other languages than Spanish using the Enchant Python module\footnote{Available at {\tt https://pypi.org/project/pyenchant/}, August 2020.}. Finally, to discard highly similar tweets, we implemented a similarity detector based on the Jaccard distance with a threshold of 0.75\%, inspired by state-of-the-art spam detection techniques \cite{Harsule2016,Bajaj2017,Temma2019}. 
 
 \item Hashtag and mention splitting. We decompose hashtags in words with a splitter that uses our own lexica \cite{Garcia-Mendez2018,Garcia-Mendez2019} and the Spanish frequency reference corpus ({\tt CREA}) by {\it Real Academia Espa\~nola de la Lengua}\footnote{Available at {\tt http://corpus.rae.es/lfrecuencias.html}, August 2020.}. As shown in the example in Table \ref{tab:beforeafterpreproOriginal}, the splitter divides the term {\em acuerdocomercial} successfully as {\em acuerdo comercial} `commercial agreement'.
 
 \item Spelling correction. We correct words with spelling mistakes by replacing them by the most likely candidate through our own word  algorithm that uses {\tt CREA} as well as the Enchant Pytdistancehon module. In the example in Table \ref{tab:beforeafterpreproOriginal}, the word {\em precausion} is corrected and replaced by {\em precaución} `caution'.
 
 \item Stock market asset, mention and hashtag detection and removal. We use regular expressions to detect capitalised letters and identify representative symbols such as \$, @ and \#. Once identified, all these elements are removed from the text of the entries.
 
 \item Stop words removal. Meaningless words such as determiners and prepositions\footnote{Available at {\tt https://www.ranks.nl/stopwords/spanish}, August 2020.} are removed from the text. We also remove {\sc url}s and retweet ({\sc rt}) tags. We consider days of the week and months of the year as stop words. However, we keep elements such as {\em no} `not', {\em sí} `yes', {\em muy} `very' and {\em poco} `few', since they help to interpret the evolution of the assets.
 
 \item Quantitative data and laughter replacement. We substitute quantitative data (numbers and percentages) by tags representing their sign, that is, `+' and `-' for positive and negative figures, respectively. We decided to use these symbolic representations rather than the amounts themselves because they are more clearly related to stock markets' upward and downward movements. Laughing onomatopoeic expressions such as {\em haha} and {\em hehe}, which are typically composed of the characters {\em j} or {\em h} interspersed with the vowels {\em a}, {\em e} and {\em i} in any quantity, are replaced by the {\tt LAUGH} tag.
 
 \item Text lemmatisation. The content of the tweets is split into tokens (words), which are independently checked in the Hunspell dictionary\footnote{Available at {\tt https://hunspell.github.io/}, August 2020.}. Finally, the words are lemmatised using the Freeling {\sc nlp} tool\footnote{Available at {\tt http://nlp.lsi.upc.edu/freeling/node/1}, August 2020.}.

\end{itemize}

\begin{table*}[!htbp]
\centering
\caption{Examples of tweets before and after processing.}
\begin{tabular}{cc}
\hline
& \textbf{Tweet} \\ \toprule
Before & \begin{tabular}[c]{@{}p{14cm}}{\em \$VIX continúa subiendo... Mucho cuidado señores! Mantengan precausion apenas estamos a inicio de semana, observemos!} \\`\$VIX continues to rise... Be very careful gentlemen! Maintain {\it precausion} we are just at the beginning of the week, let's watch!'\\\\

{\em En lo que va de año:\$IBEX: +2.26\% \$STXE (Eurostoxx Telecom): -2.9\% \$TEF: -16.75\% https://t.co/uHXa13LqoH}\\ `Year to date: \$IBEX: +2.26 \% \$STXE (Eurostoxx Telecom): -2.9 \% \$TEF: -16.75 \% https://t.co/uHXa13LqoH'\\\\

{\em \#dow \#sp500 \#nasdaq \#dax un poco de humor. Jajaja https://t.co/ecKayZEcw6}\\ `\#dow \#sp500 \#nasdaq \#dax a little humour. Hahaha https://t.co/ecKayZEcw6'\\\\

{\em El \#Ibex35 trunca la racha bajista animado por la firma del \#acuerdocomercial https://t.co/TPdOeBN3Jv vía @valenciaplaza}\\ `\#Ibex35 truncates the downward streak encouraged by the signature of the \#commercialagreement https://t.co/TPdOeBN3Jv via @valenciaplaza'
\end{tabular}
\\ \hline
After & \begin{tabular}[c]{@{}p{14cm}}{\em continuar subir cuidado señor ! mantener precaución apenas inicio observar !}\\ `continue rise careful gentleman! maintain caution just begin watch!'\\\\

{\em ir año + - -}\\ `go year + - -'\\\\

{\em poco humor LAUGH}\\ `little humour LAUGH'\\\\

{\em truncar racha bajista animar firma acuerdo comercial vía valencia plaza}\\ `truncate downward streak encourage signature commercial agreement via valencia plaza'
\end{tabular}
\\ \hline
\end{tabular}
\label{tab:beforeafterpreproOriginal}
\end{table*}

\subsection{Classification module}

\subsubsection{Machine Learning model: feature processing}\label{sec:feature_list}

The features we selected to extract valuable information from our dataset and build our model are:

\begin{itemize}

 \item Char-grams, word tokens and word-grams. Char-grams are sequences of adjacent characters in a text (note that spaces must also be considered in this case). Word tokens represent character grams only from the text inside word boundaries, whereas word \textit{n}-grams are sequences of adjacent words in a text, among which we also include the textual representation of the emojis in the tweets. To generate char \textit{n}-grams, word tokens and word \textit{n}-grams we combined {\it CountVectorizer}\footnote{Available at {\tt https://scikit-learn.org/stable/
 modules/generated/sklearn.feature\_extraction
 .text.CountVectorizer.html}, July 2020.} with {\it GridSearchCV}\footnote{Available at {\tt https://scikit-learn.org/stable/
 modules/generated/sklearn.model\_selection
 .GridSearchCV.html}, August 2020.}, both from the Scikit-Learn Python library, using the parameters in Listing \ref{configuration_parameters}. {\it GridSearchCV} performs an exhaustive search over specified parameter ranges for an estimator. As a result, we obtained the optimal parameters max\_df = 0.5, min\_df = 0.001 and ngram\_range = (1,7) in our scenario.

 \item Frequency counters. These features count the number of hashtags, negative and positive amounts and percentages, exclamation marks, interrogation marks and adverbs in the content of the tweets.

 \item Dictionary features. We use our polarity and emoji lexicon\footnote{\label{note1}Available at {\tt https://www.gti.uvigo.es/index.php/en
 /resources/8-lexicon-of-polarity-and-list-of-emojis
 -by-polarity-and-emotion-for-application
 -in-the-financial-field}, August 2020.}, for negative and positive sentiments (specifically for characterisation of general negative and positive financial emotions), negative awareness, positive statement and opportunity financial emotions; and an external emotion lexicon\footnote{Available at {\tt http://www.cic.ipn.mx/$\sim$sidorov/SEL.txt}, August 2020.}, for sadness and happiness. We count the words and emojis in the tweets corresponding to each sentiment and emotion category of those lexica.
 
 \begin{lstlisting}[frame=single,caption={Configuration for the generation of $n$-grams.},basicstyle=\small, label={configuration_parameters}]
max_df: (0.3,0.35,0.4,0.5,0.7,0.8,1)
min_df: (0,0.001,0.005,0.008,0.01)
ngram_range: ((1,1),(1,2),(1,3),(1,4),
             (1,4),(1,5),(1,6),(1,7))
max_features: (10000,20000,30000,None)
\end{lstlisting}
 
 \item Temporal features. These features, based on the discursive analysis in the preprocessing stage, are relevant because speculative tweets (such as those expressing caution or opportunities) tend to use certain verbal tenses (such as conditional tenses) in conjunction with particular expressions like {\em alto riesgo} `high risk' and {\em ocasión inmejorable} `unbeatable occasion'. The temporal features of a tweet count the verbs in past, present, future, and conditional tenses.

\end{itemize}

Table \ref{tab:features} summarises the features to build our Machine Learning model for financial opportunity detection.

\begin{table*}[ht!]
\centering
\small
\caption{\label{tab:features} Input features in the Machine Learning model.}
\begin{tabular}{lll}
\hline
\bf Type & \bf Feature name & \bf Description \\\toprule

Textual & CHAR\_GRAMS & \begin{tabular}[c]{@{}p{8.5cm}@{}} Sequence of character $n$-grams\end{tabular}\\

 & WORD\_TOKENS & \begin{tabular}[c]{@{}p{8.5cm}@{}} Sequence of character $n$-grams only from text inside word boundaries\end{tabular}\\

 & WORD\_GRAMS & \begin{tabular}[c]{@{}p{8.5cm}@{}} Sequence of word $n$-grams\end{tabular}\\ \hline

Numerical & NEG\_NUM & \begin{tabular}[c]{@{}p{8.5cm}@{}} Amount of negative numerical values\end{tabular}\\

 & POS\_NUM & \begin{tabular}[c]{@{}p{8.5cm}@{}} Amount of positive numerical values\end{tabular}\\
 
 & NEG\_PERC & \begin{tabular}[c]{@{}p{8.5cm}@{}} Amount of negative percentages\end{tabular}\\
 
 & POS\_PERC & \begin{tabular}[c]{@{}p{8.5cm}@{}} Amount of positive percentages\end{tabular}\\

 & HASHTAG & \begin{tabular}[c]{@{}p{8.5cm}@{}} Amount of hashtags\end{tabular}\\
 
 & EXCLAMATION & \begin{tabular}[c]{@{}p{8.5cm}@{}} Amount of exclamation marks\end{tabular}\\
 
 & INTERROGATION & \begin{tabular}[c]{@{}p{8.5cm}@{}} Amount of interrogation marks\end{tabular}\\
 
 & ADVERBS & \begin{tabular}[c]{@{}p{8.5cm}@{}} Amount of adverbs\end{tabular}\\
 
 & NEG\_POLARITY & \begin{tabular}[c]{@{}p{8.5cm}@{}} Amount of words with negative polarities\end{tabular}\\
 
 & NEU\_POLARITY & \begin{tabular}[c]{@{}p{8.5cm}@{}} Amount of words with neutral polarities\end{tabular}\\
 
 & POS\_POLARITY & \begin{tabular}[c]{@{}p{8.5cm}@{}} Amount of words with positive polarities\end{tabular}\\
 
 & SADNESS\_EMOTION & \begin{tabular}[c]{@{}p{8.5cm}@{}} Amount of words that express sadness\end{tabular}\\

 & HAPPINESS\_EMOTION & \begin{tabular}[c]{@{}p{8.5cm}@{}} Amount of words that express happiness\end{tabular}\\
 
 & NEG\_EMOJI & \begin{tabular}[c]{@{}p{8.5cm}@{}} Amount of emojis that express negativity\end{tabular}\\
 
 & POS\_EMOJI & \begin{tabular}[c]{@{}p{8.5cm}@{}} Amount of emojis that express positivity\end{tabular}\\
 
 & NEG\_AWARENESS\_EMOJI & \begin{tabular}[c]{@{}p{8.5cm}@{}} Amount of emojis that express negative awareness about assets\end{tabular}\\
 
 & POS\_STATEMENT\_EMOJI & \begin{tabular}[c]{@{}p{8.5cm}@{}} Amount of emojis that express positive statements about assets\end{tabular}\\
 
 & OPPORTUNITY\_EMOJI & \begin{tabular}[c]{@{}p{8.5cm}@{}} Amount of emojis that express opportunities about assets\end{tabular}\\ \hline
 
Temporal & PAST & \begin{tabular}[c]{@{}p{8.5cm}@{}} Amount of verbs in past tense\end{tabular}\\

 & PRESENT & \begin{tabular}[c]{@{}p{8.5cm}@{}} Amount of verbs in present tense\end{tabular}\\

 & FUTURE & \begin{tabular}[c]{@{}p{8.5cm}@{}} Amount of verbs in future tense\end{tabular}\\

 & CONDITIONAL & \begin{tabular}[c]{@{}p{8.5cm}@{}} Amount of verbs in conditional tense\end{tabular}\\

\bottomrule
\end{tabular}
\end{table*}

\subsubsection{Emotion analysis: three-layer stacked system}\label{sec:classification_module}

\begin{figure*}[!htbp]
 \centering
 \includegraphics[scale=0.45]{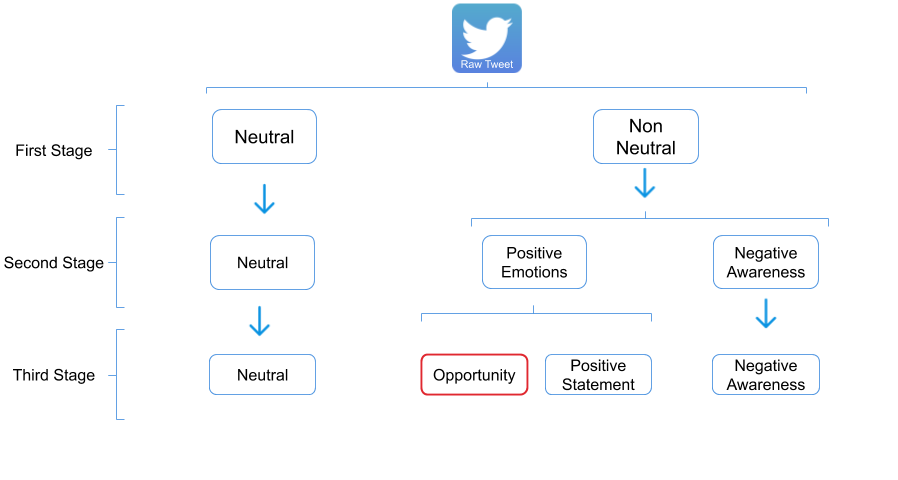}
 \caption{Flow diagram of our stacked classifier.}
 \label{fig:classifiercascadepolarity}
\end{figure*}

Figure \ref{fig:classifiercascadepolarity} shows the flow diagram of the stacked system. It is composed of three stages: a first stage that distinguishes between neutral and non-neutral entries, a second stage to distinguish between positive and negative emotions and a last stage to extract opportunities from positive statements. In each of these stages, we applied a decision depth threshold to maximise opportunity detection, because in financial analysis a high precision in key categories is preferable over obtaining a large amount of positives \cite{Jurgovsky2018}. 

The system includes implementations of Gradient Descent ({\sc gd}), Decision Tree ({\sc dt}), Random Forest ({\sc rf}) and Support Vector Classification ({\sc svc}) algorithms from the Scikit-Learn Python library\footnote{Available at {\tt https://scikit-learn.org/stable/
supervised\_learning.html\#supervised-learning}, August 2020.}. Note that the size of our training datasets did not justify the application of {\sc dl} techniques \cite{Papernot2015,Keshari2020}. In this regard, works such as \cite{Abdul-Mageed2017,Santamaria-Granados2019} are representative state of the art examples of {\sc dl} techniques for {\sc ea} where the datasets are much larger than ours.

Accordingly, we must characterise the level of confidence (depth level) of emotion prediction precision. This can be formalised with a non-parametric approach based on isotonic regression \cite{Luss2014}. This approach fits data with a non-decreasing function defining a partial or total ordering, by minimising quadratic expression \eqref{equation:depth1} subject to \eqref{equation:depth2}.

\begin{equation}
\label{equation:depth1}
\sum_{i}\alpha_{i}(y_{i}-\hat{y}_{i})^2
\end{equation}

\begin{equation}
\label{equation:depth2}
\hat{y}_{min}=\hat{y}_{1}\leq\hat{y}_{2}...\leq\hat{y}_{n}=\hat{y}_{max}
\end{equation}

In expression \eqref{equation:depth1}, each variable $\alpha_{i}$ is a strictly positive weight (1 by default), $y_{i}$ $\in$ $N$ is the true category of the $i$-th tweet and $\hat{y}_{i}$ $\in$ $N$ its predicted category (so that $y_{i}$ and $\hat{y}_{i}$ are taken to a numerical space that allows defining a partial or total ordering).

The level of confidence in the prediction depends on the probabilities of correct class assignment. Logically, it is possible to select subsets of vectors in the training set for which such probabilities are higher, and thus to trade accuracy by precision.

\section{Results and discussion}\label{sec:preliminaryTest}

All the experiments were executed on a computer with the following specifications:
\begin{itemize}
 \item Operating System: Ubuntu 18.04.2 LTS 64 bits
 \item Processor: Intel\@Core i9-9900K 3.60 GHz 
 \item RAM: 32GB DDR4 
 \item Disk: 500 Gb (7200 rpm SATA) + 256 GB SSD
\end{itemize}

\subsection{Dataset}\label{sec:dataset_information}
TxStockData S.L., a fintech company, provided us with a dataset composed of 6,000 tweets (similar in size to those in other state-of-the-art studies \cite{Chatzis2018,Al-Smadi2019,Simester2019,Tuke2020}). 
The dataset was gathered from 14th May 2019 to 3rd February 2020. Its entries were manually annotated with financial emotion tags by five experts in finance and {\sc nlp}. At the end, the emotion tag of each entry in the dataset was computed by majority voting (in case of tie, it was selected randomly). To ensure the quality of the data, we discarded repeated and spam entries, yielding approximately 5,000 valid tweets \footnote{We will make this dataset available to other researchers on request.}. Each entry has the following structure:
\begin{itemize}
 \item ID: a unique numerical identifier.
 \item Tweet: original text of the Tweet.
 \item Ticker: stock market assets mentioned in the text.
 \item Emotion: emotion label.
\end{itemize}

Table \ref{tab:datasetCharacteristic} shows the distribution of the entries in the experimental dataset by emotion category. 

\begin{table}[!htbp]
\centering
\caption{Distribution of samples in the dataset by category.}
\begin{tabular}{cc}
\toprule \textbf{Emotion} & \textbf{Number of entries} \\ \toprule
Opportunity ($P^+$) & 1,198\\
Positive statement ($S^+$) & 669 \\ 
Neutral ($N$) & 1,289 \\ 
Negative awareness ($A^-$) & 1,803 \\\hline
Total & 4,959 \\ \bottomrule
\end{tabular}
\label{tab:datasetCharacteristic}
\end{table}

\subsection{Definition of performance metrics}
\label{sec:preliminary_tests}

The practical goal of our system is to present to the user the content of the tweets that are marked as opportunities. Therefore, as previously said, we are interested in maximising precision for the target financial emotion opportunity. We defined two auxiliary tolerance metrics for those output decisions that do not correspond to true opportunities but will not necessarily discourage an operator, inspired by works like \cite{Jurgovsky2018}.

We explain these tolerance metrics with the confusion matrix in Table \ref{tab:confusionmatrix2}. Notation $X_{Y}$ represents the number of tweets of emotion category $X$ that are classified into category $Y$.

\begin{table}[!htbp]
\centering
\caption{Confusion matrix for tolerance calculations.}
\begin{tabular}{lc}
 & \\ 
\multicolumn{1}{l}{} & 
$\begin{pmatrix} 
S^+_{S^+} & S^+_{P^+} & S^+_{N} & S^+_{A^-}\\
P^+_{S^+} & P^+_{P^+} & P^+_{N} & P^+_{A^-}\\
N_{S^+} & N_{P^+} & N_{N} & N_{A^-}\\
A^-_{S^+} & A^-_{P^+} & A^-_{N} & A^-_{A^-}\\
\end{pmatrix}$ \\ 
\end{tabular}
\label{tab:confusionmatrix2}
\end{table}

From this confusion matrix we define tolerances $\uptau_1$ and $\uptau_2$ as follows:

\begin{equation}
 \uptau_1 = \frac{S^+_{P^+} + P^+_{P^+}}{S^+_{P^+} + P^+_{P^+} + N_{P^+} + A^-_{P^+}}
\end{equation}
\begin{equation}\label{eq:tol2}
 \uptau_2 = \frac{S^+_{P^+} + P^+_{P^+} + N_{P^+}}{S^+_{P^+} + P^+_{P^+} + N_{P^+} + A^-_{P^+}}
\end{equation}

In both cases the denominator is the sum of all decisions marked as opportunities. Tolerance $\uptau_1$ grows with $S^+_{P+}$, that is, when the user is presented positive statements as opportunities. Strictly these decisions are errors, but they have minimal impact on user trust if $P^+_{P+}>>S^+_{P+}$. Tolerance $\uptau_2$ is the inverse of the probability of presenting negative awareness results as opportunities, $A^-_{P^+}$, which would strongly discourage an operator, that is $\uptau_2=1- \frac{A^-_{P^+}}{S^+_{P^+} + P^+_{P^+} + N_{P^+} + A^-_{P^+}}$ (where $N_{P^+}$ corresponds to neutral tweets from a financial perspective). Therefore, our declared goal is first and foremost a reasonably high precision in the detection of opportunities, for high values of $\uptau_1$ and, secondarily, of $\uptau_2$. 

\subsection{Results}\label{sec:numericaltest}

In this section, we evaluate our system to detect financial opportunities. First we analyse the contributions the features of the Machine Learning model (see Table \ref{tab:features}) using a basic single-layer classifier of opportunities vs. rest of emotions (Section \ref{sec:featureimportance}). Then we present the precision and tolerance results of a two-layer (neutral vs. non-neutral / opportunities vs. rest of emotions) stacked classifier, without (Section \ref{sec:twolayoutbehaviour}) and with (Section \ref{sec:keywordsstageresults}) decision depth thresholds. Finally, we present our final three-layer stacked classifier (which adds a final opportunities vs. positive statements layer) with decision depth thresholds, culminating our incremental design (Section \ref{sec:threelayoutbehaviourwithdepth}). All results presented in this section were computed by applying 10-fold cross-validation.

Our model, with all characteristics, has $\sim$400,000 $n$-gram features. This is computationally intractable. Consequently, we applied an attribute selector to extract the most relevant features with a maximum decrease of 0.5\% in precision. Specifically, we chose the {\em SelectPercentile}\footnote{\label{note2}Available at {\tt https://scikit-learn.org/stable/modules
/feature\_selection.html}, August 2020.} method from the Scikit-Learn Python library, as it outperformed other alternatives ({\em SelectFromModel}, {\em SelectKBest}, and {\em RFECV}). This method selects features according to a highest score percentile. We set a $\chi^2$ score function and an 80th percentile threshold. At the end, we kept $\sim$50,000 $n$-gram features.

\subsubsection{Numerical test 1: relevance of the features}\label{sec:featureimportance}

Firstly we evaluated the performance of a single-layer classifier that was trained to distinguish between opportunities and all the other financial emotions in Table \ref{tab:datasetCharacteristic}. Table \ref{tab:resultsnumericaltest1} shows average results for the target emotion opportunity with 10-fold cross-validation. We started with a basic feature set only containing char-grams, word tokens and word-grams. Adding all features in Section \ref{sec:feature_list} was slightly advantageous for all classifiers (up to $\sim$ 4\% of improvement) except for the {\sc svc}. Moreover, in light of the results, {\sc rf} seemed the best classifier to detect opportunities, followed by the {\sc gd} classifier.

\begin{table*}[!htbp]
\centering
\caption{\label{tab:resultsnumericaltest1}Precisions and tolerances for numerical test 1.}
\small
\begin{tabular}{ccccc}
\toprule
\bf Classifier & \bf Features & \bf Precision & $ \bf \uptau_1 $ & $\bf \uptau_2 $\\ \hline
\multirow{2}{*}{GD}
& basic set & 61.67\% & 68.72\% & 84.14\% \\
& all features & 62.69\% & 70.38\% & \bf 87.31\% \\\cline{2-5}

\multirow{2}{*}{DT}
& basic set & 37.79\% & 49.25\% & 69.77\% \\
& all features & 40.25\% & 52.57\% & 73.58\% \\\cline{2-5}

\multirow{2}{*}{SVC}
& basic set & 58.48\% & 68.75\% & 85.71\% \\
& all features & 55.97\% & 67.08\% & 84.36\% \\\cline{2-5}

\multirow{2}{*}{RF}
& basic set & 60.84\% & 67.37\% & 84.21\% \\
& all features & \bf 64.33\% & \bf 70.65\% & 86.91\% \\
\bottomrule
\end{tabular}
\end{table*}

We considered these initial results (precision, $\uptau_1$ and $\uptau_2$ up to 64.33\%, 70.65\% and 87.31\%, respectively) promising and they motivated us to design more advanced approaches. The difference between $\uptau_1$ and $\uptau_2$, which ranged between 15\% and 21\%, indicated that a significant amount of neutral entries were considered as opportunities. This is why we decided to explore the benefits of a stacked classifier with an initial stage to filter neutral entries.

\subsubsection{Numerical test 2: two-layer stacked classifier}\label{sec:twolayoutbehaviour}

In this test we evaluated a two-layer stacked classifier. The first stage was designed two distinguish between neutral and non-neutral entries and the second stage distinguished opportunities from the rest of financial emotions in Table \ref{tab:datasetCharacteristic}. In both stages we used the same type of classifier but we applied independent hyperparameter optimisation in each stage. Table \ref{tab:resultsnumericaltest2} shows the resulting precisions and tolerances.

\begin{table*}[!htbp]
\centering
\caption{\label{tab:resultsnumericaltest2}Precisions and tolerances of the two-layer stacked classifier, numerical test 2.}
\small
\begin{tabular}{ccccc}
\toprule
\bf Classifier & \bf Features & \bf Precision & $ \bf \uptau_1 $ & $\bf \uptau_2 $\\ \hline
\multirow{2}{*}{GD}
& basic set & 61.37\% & 66.95\% & 82.83\% \\
& all features & 63.52\% & 71.24\% & 85.84\% \\\cline{2-5}

\multirow{2}{*}{DT}
& basic set & 37.17\% & 49.39\% & 70.16\% \\
& all features & 39.53\% & 51.61\% & 72.99\% \\\cline{2-5}

\multirow{2}{*}{SVC}
& basic set & 62.09\% & 72.53\% & 87.91\% \\
& all features & 61.10\% & 71.84\% & 87.83\% \\\cline{2-5}

\multirow{2}{*}{RF}
& basic set & 68.67\% & 73.82\% & \bf 87.98\% \\
& all features & \bf 70.26\% & \bf 75.43\% & 87.93\% \\
\bottomrule
\end{tabular}
\end{table*}

If we compare this test with numerical test 1, there was no significant improvement with the {\sc gd} and {\sc dt} classifiers but the stacked approach produced a 5\% performance boost for the other two. More specifically, the {\sc svc} and {\sc rf} classifiers produced the best results for all metrics and both scenarios (basic set and all features, except for the precision of the {\sc svc} if compared with the {\sc gd} classifier with all features). Furthermore, even though the {\sc gd} and {\sc svc} classifiers behaved similarly with all features, the {\sc gd} classifier is far more computationally intensive. {\sc dt} precision results were considerably bad, of around 40\%. Thus we decided to discard {\sc gd} and {\sc dt} for opportunity detection and keep only {\sc svc} and {\sc rf} in further analysis.

Ultimately, even though this test revealed further improvement in precision and tolerance with the two-level stacked classifier, the highest precision (with the {\sc rf} algorithm) was still under 75\%. However, the tolerances against operator discouragement were moderately satisfactory ($\uptau_1>$75\% and $\uptau_2>$85\%).

\subsubsection{Numerical test 3: two-layer stacked classifier with decision depth}\label{sec:keywordsstageresults}

As previously stated, in businesses and investment, it is preferable to obtain less positives for key categories with high precision. Take personalised investment advertising as an example. Banks must identify customer profiles with higher success probability, since personalised commercialisation actions are rather expensive. From a practical perspective, we set the goal of detecting at least 10\% of all opportunity entries with high precision.

Accordingly, in this test we relied on the {\tt predict\_proba} setting of the Scikit-Learn Python library to set class thresholds. It allows computing the probabilities of the possible outcomes (classes in our model), thus providing levels of confidence (see Section \ref{sec:classification_module}).
When the probability of the most likely class for a vector exceeded the threshold of that class (which we term decision depth) the vector was assigned that class. Otherwise, the entry was considered neutral. This procedure was followed at both stages of numerical test 2, with the same depth value, after empirical tests. 

Table \ref{tab:resultsnumericaltest3} shows the results of the two-layer stacked classifier with decision depth. Both classifiers achieved significant precision improvements. Besides, the {\sc rf} classifier achieved a precision above 70\% in all cases for $\uptau_1$ $\sim80\%$ and $\uptau_2$ $\sim90\%$. Regardless of the fact that precision and $\uptau_1$ were still not satisfactory (they were less than 75\% and 90\%, respectively), 
the values of $\uptau_2$ for the two classifiers with all features indicated that many negative awareness entries were still classified as opportunities, which was unacceptable.

\begin{table*}[!htbp]
\centering
\caption{\label{tab:resultsnumericaltest3}Precisions and tolerances of the two-layer stacked classifier with decision depth, numerical test 3.}
\small
\begin{tabular}{cccccc}
\toprule
\bf Classifier & \bf Features & \bf Precision & $ \bf \uptau_1 $ & $\bf \uptau_2 $ & \bf Depth\\ \hline
\multirow{2}{*}{SVC}
& basic set & 61.54\% & 72.53\% & 86.81\% & 58\% \\
& all features & 66.24\% & 72.61\% & \bf 89.81\% & 62\% \\\cline{2-6}

\multirow{2}{*}{RF}
& basic set & \bf 72.93\% & 77.35\% & 87.85\% & 53\% \\
& all features & 72.61\% & \bf 79.62\% & 88.54\% & 54\% \\
\bottomrule
\end{tabular}
\end{table*}

\subsubsection{Numerical test 4: three-layer stacked classifier with decision depth}\label{sec:threelayoutbehaviourwithdepth}

All points considered, to meet our goals we finally applied the three-layer stacked classifier framework (see Section \ref{sec:classification_module}) with decision depth thresholds where the third stage distinguishes opportunities from positive statement entries. 

We relied once again on the {\tt predict\_proba} setting of the Scikit-Learn Python library to set the depth threshold. As in the previous numerical test we used the same classifier and depth threshold in all layers. Table \ref{tab:resultsnumericaltest4} shows the results of this test. Note the significant improvements by the {\sc svc} and {\sc rf} classifiers for all metrics. With the {\sc rf} classifier we obtained precision values above 80\%, $\uptau_1$ above 90\% and $\uptau_2$ above 95\%, meeting our initial requirements. Less than 5\% of the negative awareness entries were misclassified as opportunities, probably due to high ambiguity and, thus, annotation errors.

\begin{table*}[!htbp]
\centering
\caption{\label{tab:resultsnumericaltest4}Precisions and tolerances of the three-layer stacked classifier with decision depth, numerical test 4.}
\small
\begin{tabular}{cccccc}
\toprule
\bf Classifier & \bf Features & \bf Precision & $ \bf \uptau_1 $ & $\bf \uptau_2 $ & \bf Depth\\ \hline
\multirow{2}{*}{SVC}
& basic set & 76.67\% & 80.67\% & 94.00\% & 82\% \\
& all features & 77.16\% & 82.72\% & 95.06\% & 82\% \\\cline{2-6}

\multirow{2}{*}{RF}
& basic set & 82.07\% & 88.66\% & 94.48\% & 75\% \\
& all features & \bf 82.84\% & \bf 90.03\% & \bf 95.52\% & 75\% \\
\bottomrule
\end{tabular}
\end{table*}

\subsection{Exploiting the results}\label{sec:realscenario}

For the sake of clarity and a handy comparison of the different numerical tests, Table \ref{tab:comparison} shows the differences in precision, $\uptau_1$ and $\uptau_2$ between numerical tests 1 and 4. The improvement achieved with {\sc svc} and {\sc rf} classifiers is noteworthy, supporting the benefits of a stacking approach for financial opportunity detection. 
The performance boost for {\sc rf}, our final choice, was almost 20\% for precision and $\uptau_1 $. In addition to meeting our performance goals, we succeeded to identify as many opportunity entries with high precision as our application demanded, $9.27\%$ and $10.43\%$ on average with {\sc rf} and {\sc svc} respectively, with 10-fold cross-validation. 

\begin{table*}[!htbp]
\centering
\caption{\label{tab:comparison}Improvements in precisions and tolerances between numerical tests 1 and 4.}
\small
\begin{tabular}{ccccc}
\toprule
\bf Classifier & \bf Features & $ \bf \Delta$ \bf Precision & $ \bf \Delta \uptau_1 $ & $\bf \Delta \uptau_2 $\\ \hline
\multirow{2}{*}{SVC}
& basic set & 18.19\% & 11.92\% & 8.29\% \\
& all features & 21.19\% & 15.64\% & 10.70\% \\\cline{2-5}

\multirow{2}{*}{RF}
& basic set & 21.23\% & 21.29\% & 10.27\% \\
& all features & 18.51\% & 19.38\% & 8.61\% \\
\bottomrule
\end{tabular}
\end{table*}

\begin{figure*}[!htbp]
 \centering
 \includegraphics[scale=0.19]{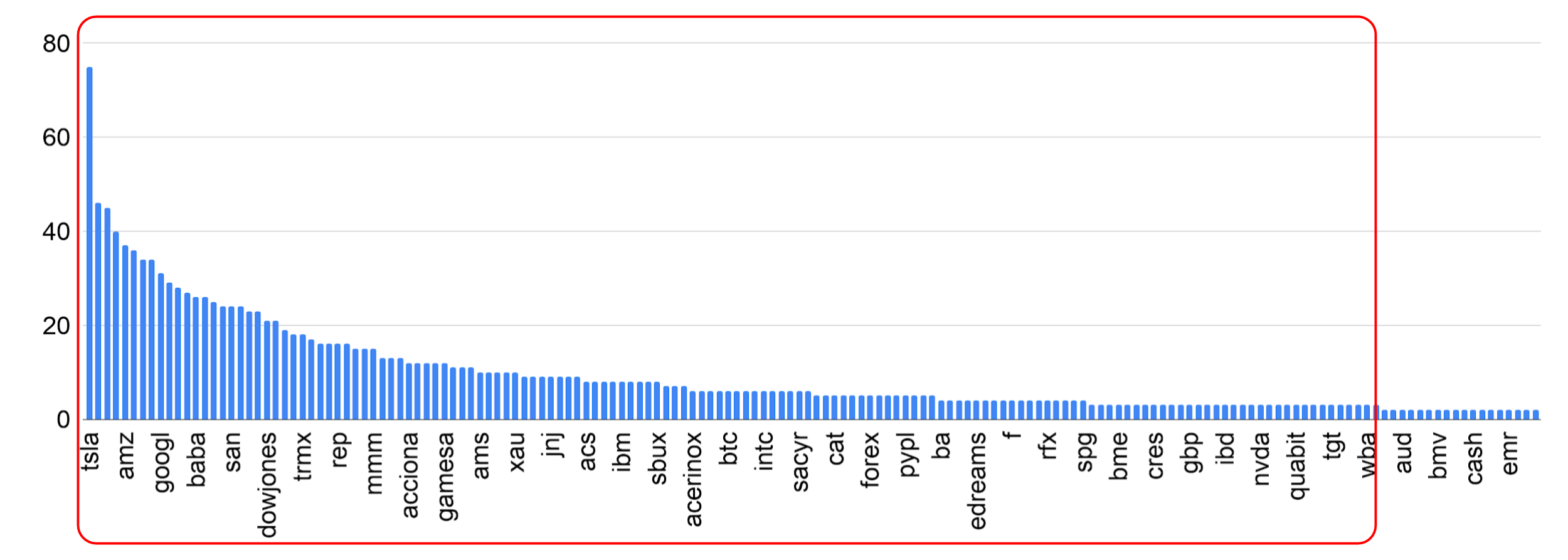}
 \caption{Histogram of ticker mentions in opportunities.}
 \label{fig:activosmas3}
\end{figure*}

Moreover, Figure \ref{fig:activosmas3} represents the histogram of the mentions to assets in opportunity tweets. By examining the specific assets that were mentioned between 14 May 2019 and 3 February 2020 in detected opportunities in the testing sets in numerical experiment 4, we observed that those assets corresponded to $\sim80\%$ of the mentions to the most interesting stock actives (upper quartile of all mentions in opportunities in the testing subsets, marked in red in Figure \ref{fig:activosmas3}. For clarity, only some representative well-known tickers are shown in the $x$ axis).

\begin{figure*}[!htbp]
 \centering
 \includegraphics[scale=0.14]{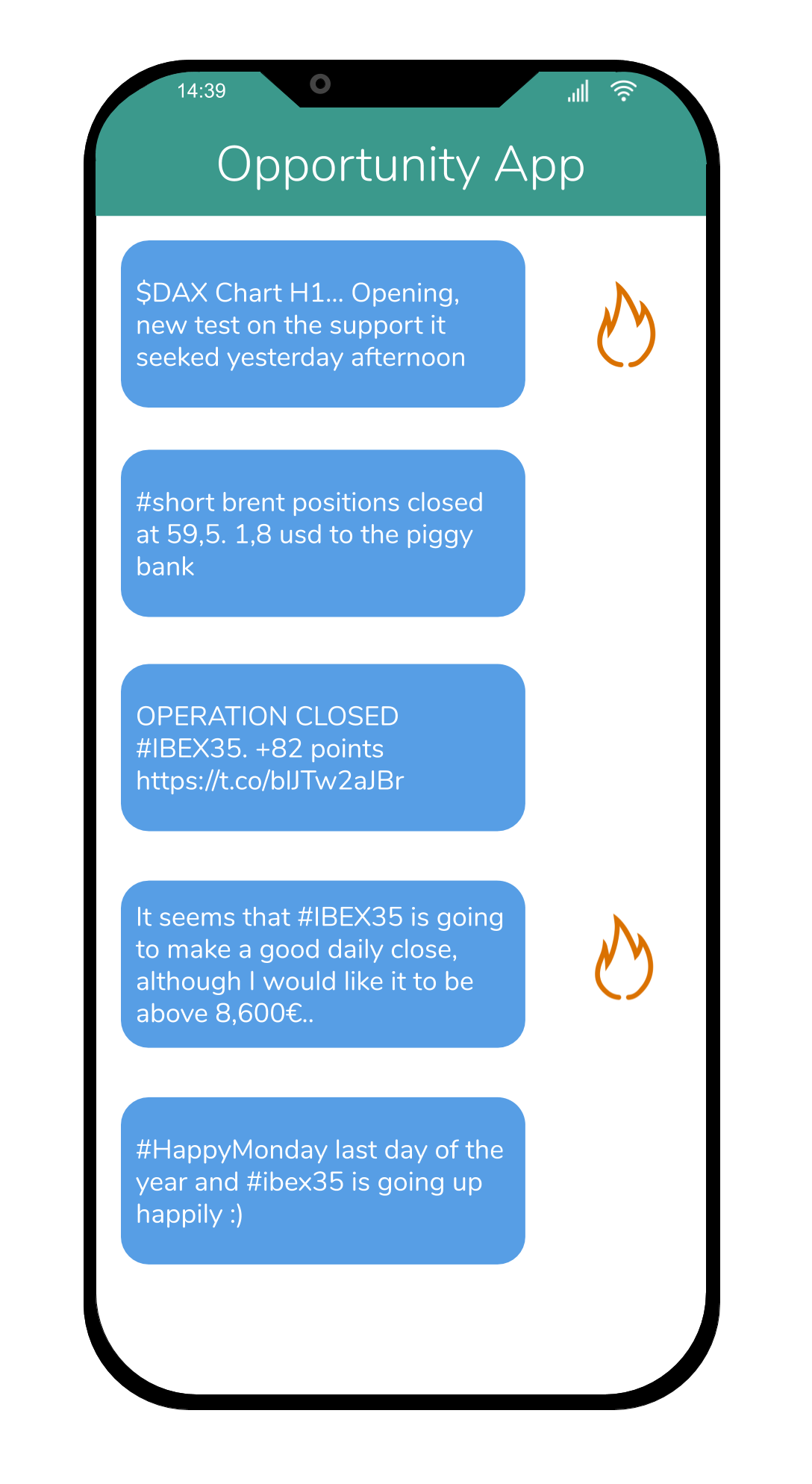}
 \caption{Possible integration of our system in a mobile app.}
 \label{fig:smartphone_app}
\end{figure*}

Therefore, a possible way to exploit these results would be listing in a scroll section of the user dashboard the tweets we pick as opportunities with the three-level stacked classifier. Then we could highlight the tweets mentioning assets that are detected following our approach with decision depth. Figure \ref{fig:smartphone_app} shows an example with real tweets as classified by our system. Symbol \includegraphics[height=0.15in]{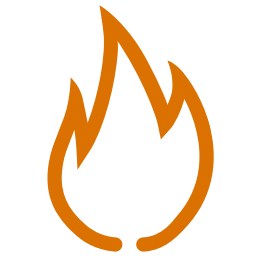} is displayed next to the tweets that contain opportunities that are detected with high precision. 

\section{Conclusions}\label{sec:conclusions}

Motivated by the strong influence of social media in users' decisions in global markets, we propose a novel three-layer stacked system to detect financial opportunities in tweets. Our solution has been designed to extract such tweets with high precision to present their content on personal finance management dashboards. It exploits sophisticated linguistic features, such as polarity and emotion dictionaries and temporal identification by discursive analysis, in its {\sc ml} model. It yields satisfactory and competitive market-level performance in financial opportunity detection.

Experimental results, including two ad-hoc tolerance metrics focusing on data quality from an operator perspective, demonstrate that our three-layer stacked system with decision probability threshold (decision depth) succeeds to achieve our goal. In particular, using a {\sc rf} algorithm with all features in the model, the system attains $\sim 83$\% financial opportunity precision with detection tolerances $\uptau_1\sim 90$\% and $\uptau_2\sim 96$\%. Given the valuable up-to-date information in micro-blogging platforms, these promising results endorse the usability of our system to support investors' decision making.

As future work and to further expand the contributions of our research, the emotion classifiers will be improved with domain-specific tweet filters and quantitative (objective) numerical information from stock exchange prices. In addition to opportunities, this will allow analysing positive statement and negative awareness emotions in stock market reports. Moreover, extensions for a multilingual version of our framework to cover French and English will be provided.

\newpage

\bibliography{bibliography.bib}{}

\begin{thebibliography}{10}
\providecommand{\url}[1]{#1}
\csname url@samestyle\endcsname
\providecommand{\newblock}{\relax}
\providecommand{\bibinfo}[2]{#2}
\providecommand{\BIBentrySTDinterwordspacing}{\spaceskip=0pt\relax}
\providecommand{\BIBentryALTinterwordstretchfactor}{4}
\providecommand{\BIBentryALTinterwordspacing}{\spaceskip=\fontdimen2\font plus
\BIBentryALTinterwordstretchfactor\fontdimen3\font minus
  \fontdimen4\font\relax}
\providecommand{\BIBforeignlanguage}[2]{{%
\expandafter\ifx\csname l@#1\endcsname\relax
\typeout{** WARNING: IEEEtran.bst: No hyphenation pattern has been}%
\typeout{** loaded for the language `#1'. Using the pattern for}%
\typeout{** the default language instead.}%
\else
\language=\csname l@#1\endcsname
\fi
#2}}
\providecommand{\BIBdecl}{\relax}
\BIBdecl

\bibitem{Li2018}
T.~Li, J.~van Dalen, and P.~J. van Rees, ``{More than just Noise? Examining the
  Information Content of Stock Microblogs on Financial Markets},''
  \emph{Journal of Information Technology}, vol.~33, no.~1, pp. 50--69, mar
  2018.

\bibitem{Li2020}
X.~Li, P.~Wu, and W.~Wang, ``{Incorporating stock prices and news sentiments
  for stock market prediction: A case of Hong Kong},'' \emph{Information
  Processing {\&} Management}, p. 102212, feb 2020.

\bibitem{Bello-Orgaz2020}
G.~Bello-Orgaz, R.~M. Mesas, C.~Zarco, V.~Rodriguez, O.~Cord{\'{o}}n, and
  D.~Camacho, ``{Marketing analysis of wineries using social collective
  behavior from users' temporal activity on Twitter},'' \emph{Information
  Processing {\&} Management}, p. 102220, feb 2020.

\bibitem{Meire2017}
M.~Meire, M.~Ballings, and D.~{Van den Poel}, ``{The added value of social
  media data in B2B customer acquisition systems: A real-life experiment},''
  \emph{Decision Support Systems}, vol. 104, pp. 26--37, dec 2017.

\bibitem{Pai2018}
P.-F. Pai and C.-H. Liu, ``{Predicting Vehicle Sales by Sentiment Analysis of
  Twitter Data and Stock Market Values},'' \emph{IEEE Access}, vol.~6, pp.
  57\,655--57\,662, 2018.

\bibitem{Yuan2018}
H.~Yuan, W.~Xu, Q.~Li, and R.~Lau, ``{Topic sentiment mining for sales
  performance prediction in e-commerce},'' \emph{Annals of Operations
  Research}, vol. 270, no. 1-2, pp. 553--576, nov 2018.

\bibitem{Mai2018}
F.~Mai, Z.~Shan, Q.~Bai, X.~S. Wang, and R.~H. Chiang, ``{How Does Social Media
  Impact Bitcoin Value? A Test of the Silent Majority Hypothesis},''
  \emph{Journal of Management Information Systems}, vol.~35, no.~1, pp. 19--52,
  jan 2018.

\bibitem{Sun2020}
Y.~Sun, X.~Liu, G.~Chen, Y.~Hao, and Z.~J. Zhang, ``{How mood affects the stock
  market: Empirical evidence from microblogs},'' \emph{Information {\&}
  Management}, vol.~57, no.~5, pp. 103--181, jul 2020.

\bibitem{Enke2005}
D.~Enke and S.~Thawornwong, ``{The Use of Data Mining and Neural Networks for
  Forecasting Stock Market Returns},'' \emph{Expert Systems with Applications},
  vol.~29, no.~4, pp. 927--940, nov 2005.

\bibitem{Gerber2014}
M.~S. Gerber, ``{Predicting Crime Using Twitter and Kernel Density
  Estimation},'' \emph{Decision Support Systems}, vol.~61, pp. 115--125, 2014.

\bibitem{Reece2017}
A.~G. Reece, A.~J. Reagan, K.~L.~M. Lix, P.~S. Dodds, C.~M. Danforth, and E.~J.
  Langer, ``{Forecasting the onset and course of mental illness with Twitter
  data},'' \emph{Scientific Reports}, vol.~7, no.~1, pp. 1--11, dec 2017.

\bibitem{Zahra2020}
K.~Zahra, M.~Imran, and F.~O. Ostermann, ``{Automatic identification of
  eyewitness messages on Twitter during disasters},'' \emph{Information
  Processing {\&} Management}, vol.~57, no.~1, pp. 102--107, 2020.

\bibitem{Oliveira2017}
N.~Oliveira, P.~Cortez, and N.~Areal, ``{The Impact of Microblogging Data for
  Stock Market Prediction: Using Twitter to Predict Returns, Volatility,
  Trading Volume and Survey Sentiment Indices},'' \emph{Expert Systems with
  Applications}, vol.~73, pp. 125--144, 2017.

\bibitem{Nofer2015}
M.~Nofer and O.~Hinz, ``{Using Twitter to Predict the Stock Market},''
  \emph{Business {\&} Information Systems Engineering}, vol.~57, no.~4, pp.
  229--242, 2015.

\bibitem{Dimpfl2016}
T.~Dimpfl and S.~Jank, ``{Can Internet Search Queries Help to Predict Stock
  Market Volatility?}'' \emph{European Financial Management}, vol.~22, no.~2,
  pp. 171--192, 2016.

\bibitem{Zhong2017}
X.~Zhong and D.~Enke, ``{Forecasting daily stock market return using
  dimensionality reduction},'' \emph{Expert Systems with Applications},
  vol.~67, pp. 126--139, jan 2017.

\bibitem{Zhang2018}
X.~Zhang, Y.~Zhang, S.~Wang, Y.~Yao, B.~Fang, and P.~S. Yu, ``{Improving stock
  market prediction via heterogeneous information fusion},''
  \emph{Knowledge-Based Systems}, vol. 143, pp. 236--247, mar 2018.

\bibitem{Hoseinzade2019}
E.~Hoseinzade and S.~Haratizadeh, ``{CNNpred: CNN-based stock market prediction
  using a diverse set of variables},'' \emph{Expert Systems with Applications},
  vol. 129, pp. 273--285, sep 2019.

\bibitem{Sun2014}
F.~Sun, A.~Belatreche, S.~Coleman, T.~M. McGinnity, and Y.~Li,
  ``{Pre-processing Online Financial Text for Sentiment Classification: A
  Natural Language Processing Approach},'' in \emph{Proceedings of the IEEE
  Conference on Computational Intelligence for Financial Engineering {\&}
  Economics}.\hskip 1em plus 0.5em minus 0.4em\relax IEEE, 2014, pp. 122--129.

\bibitem{Fisher2016}
I.~E. Fisher, M.~R. Garnsey, and M.~E. Hughes, ``{Natural Language Processing
  in Accounting, Auditing and Finance: A Synthesis of the Literature with a
  Roadmap for Future Research},'' \emph{Intelligent Systems in Accounting,
  Finance and Management}, vol.~23, no.~3, pp. 157--214, 2016.

\bibitem{Xing2018}
F.~Z. Xing, E.~Cambria, and R.~E. Welsch, ``{Natural Language Based Financial
  Forecasting: a Survey},'' \emph{Artificial Intelligence Review}, vol.~50,
  no.~1, pp. 49--73, 2018.

\bibitem{singh2016score}
K.~K. Singh and P.~Dimri, ``{Score Based Financial Forecasting Method by
  Incorporating Different Sources of Information Flow into Integrative River
  Model},'' in \emph{Proceedings of the 6th International Conference-Cloud
  System and Big Data Engineering}.\hskip 1em plus 0.5em minus 0.4em\relax
  IEEE, 2016, pp. 685--688.

\bibitem{Razi2017}
N.~I.~M. Razi, M.~Othman, and H.~Yaacob, ``{Investment Decisions Based on EEG
  Emotion Recognition},'' \emph{Advanced Science Letters}, vol.~23, no.~11, pp.
  11\,345--11\,349, 2017.

\bibitem{Plutchik2004}
R.~Plutchik, ``{The circumplex as a general model of the structure of emotions
  and personality.}'' in \emph{Circumplex models of personality and
  emotions.}\hskip 1em plus 0.5em minus 0.4em\relax American Psychological
  Association, 2004, pp. 17--45.

\bibitem{Chatzis2018}
S.~P. Chatzis, V.~Siakoulis, A.~Petropoulos, E.~Stavroulakis, and
  N.~Vlachogiannakis, ``{Forecasting stock market crisis events using Deep and
  statistical Machine Learning techniques},'' \emph{Expert Systems with
  Applications}, vol. 112, pp. 353--371, 2018.

\bibitem{Al-Smadi2019}
M.~Al-Smadi, M.~Al-Ayyoub, Y.~Jararweh, and O.~Qawasmeh, ``{Enhancing
  Aspect-Based Sentiment Analysis of Arabic Hotels' reviews using
  morphological, syntactic and semantic features},'' \emph{Information
  Processing {\&} Management}, vol.~56, no.~2, pp. 308--319, mar 2019.

\bibitem{Simester2019}
D.~Simester, A.~Timoshenko, and S.~I. Zoumpoulis, ``{Targeting Prospective
  Customers: Robustness of Machine-Learning Methods to Typical Data
  Challenges},'' \emph{Management Science}, pp. 1--43, 2019.

\bibitem{Tuke2020}
J.~Tuke, A.~Nguyen, M.~Nasim, D.~Mellor, A.~Wickramasinghe, N.~Bean, and
  L.~Mitchell, ``{Pachinko Prediction: A Bayesian method for event prediction
  from social media data},'' \emph{Information Processing {\&} Management},
  vol.~57, no.~2, p. 102147, mar 2020.

\bibitem{Rickett2016}
L.~K. Rickett, ``{Do Financial Blogs Serve an Infomediary Role in Capital
  Markets?}'' \emph{American Journal of Business}, vol.~31, no.~1, pp. 17--40,
  2016.

\bibitem{He2017}
W.~He, F.-K. Wang, and V.~Akula, ``{Managing extracted knowledge from big
  social media data for business decision making},'' \emph{Journal of Knowledge
  Management}, vol.~21, no.~2, pp. 275--294, apr 2017.

\bibitem{Alanyali2013}
M.~Alanyali, H.~S. Moat, and T.~Preis, ``{Quantifying the Relationship Between
  Financial News and the Stock Market},'' \emph{Scientific Reports}, vol.~3,
  no.~1, pp. 3578--3584, 2013.

\bibitem{Atkins2018}
A.~Atkins, M.~Niranjan, and E.~Gerding, ``{Financial News Predicts Stock Market
  Volatility Better than Close Price},'' \emph{The Journal of Finance and Data
  Science}, vol.~4, no.~2, pp. 120--137, 2018.

\bibitem{Day2016}
M.-Y. Day and C.-C. Lee, ``{Deep Learning for Financial Sentiment Analysis on
  Finance News Providers},'' in \emph{Proceedings of the International
  Conference on Advances in Social Networks Analysis and Mining}.\hskip 1em
  plus 0.5em minus 0.4em\relax IEEE, 2016, pp. 1127--1134.

\bibitem{Wang2017}
Y.~Wang, ``{Stock Market Forecasting with Financial Micro-blog Based on
  Sentiment and Time Series Analysis},'' \emph{Journal of Shanghai Jiaotong
  University}, vol.~22, no.~2, pp. 173--179, 2017.

\bibitem{Ioanas2014}
E.~Ioanăs and I.~Stoica, ``{Social Media and its Impact on Consumers
  Behavior},'' \emph{International Journal of Economic Practices and Theories},
  vol.~4, no.~2, pp. 295--303, 2014.

\bibitem{Sun2016}
A.~Sun, M.~Lachanski, and F.~J. Fabozzi, ``{Trade the Tweet: Social Media Text
  Mining and Sparse Matrix Factorization for Stock Market Prediction},''
  \emph{International Review of Financial Analysis}, vol.~48, pp. 272--281,
  2016.

\bibitem{Ming2014}
F.~Ming, F.~Wong, Z.~Liu, and M.~Chiang, ``{Stock Market Prediction from WSJ:
  Text Mining via Sparse Matrix Factorization},'' in \emph{Proceedings of the
  International Conference on Data Mining}.\hskip 1em plus 0.5em minus
  0.4em\relax IEEE, 2014, pp. 430--439.

\bibitem{Gibbs1998}
P.~Gibbs, ``{Time, Temporality and Consumer Behaviour},'' \emph{European
  Journal of Marketing}, vol.~32, no. 11/12, pp. 993--1007, 1998.

\bibitem{Forray2005}
J.~M. Forray and J.~Woodilla, ``{Artefacts of Management Academe},'' \emph{Time
  {\&} Society}, vol.~14, no. 2-3, pp. 323--339, 2005.

\bibitem{liu2015sentiment}
B.~Liu, \emph{{Sentiment Analysis: Mining Opinions, Sentiments, and
  Emotions}}.\hskip 1em plus 0.5em minus 0.4em\relax Cambridge University
  Press, 2015.

\bibitem{Derakhshan2019}
A.~Derakhshan and H.~Beigy, ``{Sentiment analysis on stock social media for
  stock price movement prediction},'' \emph{Engineering Applications of
  Artificial Intelligence}, vol.~85, pp. 569--578, oct 2019.

\bibitem{Staiano2014}
J.~Staiano and M.~Guerini, ``{Depeche Mood: a Lexicon for Emotion Analysis from
  Crowd Annotated News},'' in \emph{Proceedings of the 52nd Annual Meeting of
  the Association for Computational Linguistics}, vol.~2.\hskip 1em plus 0.5em
  minus 0.4em\relax Association for Computational Linguistics, 2014, pp.
  427--433.

\bibitem{Ge2020}
Y.~Ge, J.~Qiu, Z.~Liu, W.~Gu, and L.~Xu, ``{Beyond negative and positive:
  Exploring the effects of emotions in social media during the stock market
  crash},'' \emph{Information Processing {\&} Management}, vol.~57, no.~4, p.
  102218, jul 2020.

\bibitem{Xu2019}
G.~Xu, Y.~Meng, X.~Qiu, Z.~Yu, and X.~Wu, ``{Sentiment Analysis of Comment
  Texts Based on BiLSTM},'' \emph{IEEE Access}, vol.~7, pp. 51\,522--51\,532,
  2019.

\bibitem{Fang2018}
Y.~Fang, H.~Tan, and J.~Zhang, ``{Multi-Strategy Sentiment Analysis of Consumer
  Reviews Based on Semantic Fuzziness},'' \emph{IEEE Access}, vol.~6, pp.
  20\,625--20\,631, 2018.

\bibitem{Balahur2015SentimentTweets}
A.~Balahur and J.~M. Perea-Ortega, ``{Sentiment Analysis System Adaptation for
  Multilingual Processing: The Case of Tweets},'' \emph{Information Processing
  {\&} Management}, vol.~51, no.~4, pp. 547--556, 2015.

\bibitem{buvcar2016sentiment}
J.~Bu{\v{c}}ar, J.~Povh, and M.~{\v{Z}}nidar{\v{s}}i{\v{c}}, ``{Sentiment
  Classification of the Slovenian News Texts},'' in \emph{Proceedings of the
  9th International Conference on Computer Recognition Systems}.\hskip 1em plus
  0.5em minus 0.4em\relax Springer, 2016, pp. 777--787.

\bibitem{Zimbra2016}
D.~Zimbra, M.~Ghiassi, and S.~Lee, ``{Brand-Related Twitter Sentiment Analysis
  Using Feature Engineering and the Dynamic Architecture for Artificial Neural
  Networks},'' in \emph{Proceedings of the 49th Hawaii International Conference
  on System Sciences}.\hskip 1em plus 0.5em minus 0.4em\relax IEEE, 2016, pp.
  1930--1938.

\bibitem{Smailovic2013PredictiveApplication}
J.~Smailovi{\'{c}}, M.~Gr{\v{c}}ar, N.~Lavra{\v{c}}, and
  M.~{\v{Z}}nidar{\v{s}}i{\v{c}}, ``{Predictive Sentiment Analysis of Tweets: A
  Stock Market Application},'' in \emph{Lecture Notes in Computer
  Science}.\hskip 1em plus 0.5em minus 0.4em\relax Springer, 2013, vol. 7947,
  pp. 77--88.

\bibitem{Rout2018}
J.~K. Rout, K.-K.~R. Choo, A.~K. Dash, S.~Bakshi, S.~K. Jena, and K.~L.
  Williams, ``{A model for sentiment and emotion analysis of unstructured
  social media text},'' \emph{Electronic Commerce Research}, vol.~18, no.~1,
  pp. 181--199, mar 2018.

\bibitem{Chen2018}
C.-H. Chen, W.-P. Lee, and J.-Y. Huang, ``{Tracking and recognizing emotions in
  short text messages from online chatting services},'' \emph{Information
  Processing {\&} Management}, vol.~54, no.~6, pp. 1325--1344, nov 2018.

\bibitem{Parrott2001EmotionsReadings}
W.~G. Parrott, \emph{{Emotions in Social Psychology: Essential
  Readings}}.\hskip 1em plus 0.5em minus 0.4em\relax Psychology Press, 2001.

\bibitem{neviarouskaya2007textual}
A.~Neviarouskaya, H.~Prendinger, and M.~Ishizuka, ``{Textual Affect Sensing for
  Sociable and Expressive Online Communication},'' in \emph{Proceedings of the
  International Conference on Affective Computing and Intelligent
  Interaction}.\hskip 1em plus 0.5em minus 0.4em\relax Springer, 2007, pp.
  218--229.

\bibitem{Xu2018}
J.~Xu, Z.~Huang, M.~Shi, and M.~Jiang, ``{Emotion Detection in E-learning Using
  Expectation-Maximization Deep Spatial-Temporal Inference Network},''
  \emph{Advances in Intelligent Systems and Computing}, vol. 650, pp. 245--252,
  2018.

\bibitem{Asghar2017}
M.~Z. Asghar, A.~Khan, K.~Khan, H.~Ahmad, and I.~A. Khan, ``{COGEMO:
  Cognitive-Based Emotion Detection from Patient Generated Health Reviews},''
  \emph{Journal of Medical Imaging and Health Informatics}, vol.~7, no.~6, pp.
  1436--1444, oct 2017.

\bibitem{Bong2017}
S.~Z. Bong, K.~Wan, M.~Murugappan, N.~M. Ibrahim, Y.~Rajamanickam, and
  K.~Mohamad, ``{Implementation of wavelet packet transform and non linear
  analysis for emotion classification in stroke patient using brain signals},''
  \emph{Biomedical Signal Processing and Control}, vol.~36, pp. 102--112, jul
  2017.

\bibitem{ShamimHossain2015}
M.~S. Hossain, G.~Muhammad, B.~Song, M.~M. Hassan, A.~Alelaiwi, and A.~Alamri,
  ``{Audio–Visual Emotion-Aware Cloud Gaming Framework},'' \emph{IEEE
  Transactions on Circuits and Systems for Video Technology}, vol.~25, no.~12,
  pp. 2105--2118, 2015.

\bibitem{Sanchez-Rada2014}
J.~F. S{\'{a}}nchez-Rada, M.~Torres, C.~A. Iglesias, R.~Maestre, and
  E.~Peinado, ``{A linked data approach to sentiment and emotion analysis of
  Twitter in the financial domain},'' in \emph{Joint Proceedings of the Second
  International Workshop on Semantic Web Enterprise Adoption and Best Practice
  and Second International Workshop on Finance and Economics on the Semantic
  Web}, vol. 1240.\hskip 1em plus 0.5em minus 0.4em\relax CEUR, 2014, pp.
  51--62.

\bibitem{Duxbury2020}
D.~Duxbury, T.~G{\"{a}}rling, A.~Gamble, and V.~Klass, ``{How emotions
  influence behavior in financial markets: a conceptual analysis and
  emotion-based account of buy-sell preferences},'' \emph{The European Journal
  of Finance}, pp. 1--22, 2020.

\bibitem{Pengnate2020}
S.~F. Pengnate and F.~J. Riggins, ``{The role of emotion in P2P microfinance
  funding: A sentiment analysis approach},'' \emph{International Journal of
  Information Management}, vol.~54, p. 102138, 2020.

\bibitem{Fernandez-Gavilanes2018CreatingDescriptions}
M.~Fern{\'{a}}ndez-Gavilanes, J.~Juncal-Mart{\'{i}}nez,
  S.~Garc{\'{i}}a-M{\'{e}}ndez, E.~Costa-Montenegro, and F.~J.
  Gonz{\'{a}}lez-Casta{\~{n}}o, ``{Creating Emoji Lexica from Unsupervised
  Sentiment Analysis of their Descriptions},'' \emph{Expert Systems with
  Applications}, vol. 103, pp. 74--91, 2018.

\bibitem{alvarez2015gti}
T.~Alvarez-L{\'{o}}pez, J.~Juncal-Mart{\'{i}}nez, M.~F. Gavilanes,
  E.~Costa-Montenegro, F.~J. Gonz{\'{a}}lez-Casta\~no, H.~Cerezo-Costas, and
  D.~Celix-Salgado, ``{GTI-Gradiant at TASS 2015: A Hybrid Approach for
  Sentiment Analysis in Twitter},'' in \emph{Proceedings of the Workshop on
  Semantic Analysis at the International Conference of the Spanish Society for
  Language Processing}.\hskip 1em plus 0.5em minus 0.4em\relax CEUR, 2015, pp.
  35--40.

\bibitem{Mehmood2018}
A.~Mehmood, B.-W. On, I.~Lee, I.~Ashraf, and G.~{Sang Choi}, ``{Spam comments
  prediction using stacking with ensemble learning},'' \emph{Journal of
  Physic}, vol. 933, p. 012012, jan 2018.

\bibitem{Wang2019}
Y.~Wang, S.~Liu, S.~Li, J.~Duan, Z.~Hou, J.~Yu, and K.~Ma, ``{Stacking-Based
  Ensemble Learning of Self-Media Data for Marketing Intention Detection},''
  \emph{Future Internet}, vol.~11, no.~7, p. 155, 2019.

\bibitem{Harsule2016}
S.~R. Harsule and M.~K. Nighot, ``{N-Gram Classifier System to Filter Spam
  Messages from OSN User Wall},'' in \emph{Advances in Intelligent Systems and
  Computing}.\hskip 1em plus 0.5em minus 0.4em\relax Springer, 2016, pp.
  21--28.

\bibitem{Bajaj2017}
S.~Bajaj, N.~Garg, and S.~K. Singh, ``{A Novel User-based Spam Review
  Detection},'' \emph{Procedia Computer Science}, vol. 122, pp. 1009--1015,
  2017.

\bibitem{Temma2019}
S.~Temma, M.~Sugii, and H.~Matsuno, ``{The Document Similarity Index based on
  the Jaccard Distance for Mail Filtering},'' in \emph{Proceedings of the 34th
  International Technical Conference on Circuits/Systems, Computers and
  Communications}.\hskip 1em plus 0.5em minus 0.4em\relax IEEE, 2019, pp. 1--4.

\bibitem{Garcia-Mendez2018}
S.~Garc{\'{i}}a-M{\'{e}}ndez, M.~Fern{\'{a}}ndez-Gavilanes,
  E.~Costa-Montenegro, J.~Juncal-Mart{\'{i}}nez, and F.~J.
  Gonz{\'{a}}lez-Casta{\~{n}}o, ``{Automatic Natural Language Generation
  Applied to Alternative and Augmentative Communication for Online Video
  Content Services using SimpleNLG for Spanish},'' in \emph{Proceedings of the
  Internet of Accessible Things}.\hskip 1em plus 0.5em minus 0.4em\relax ACM,
  apr 2018, pp. 1--4.

\bibitem{Garcia-Mendez2019}
S.~Garc{\'{i}}a-M{\'{e}}ndez, M.~Fern{\'{a}}ndez-Gavilanes,
  E.~Costa-Montenegro, J.~Juncal-Mart{\'{i}}nez, and F.~{Javier
  Gonz{\'{a}}lez-Casta{\~{n}}o}, ``{A library for automatic Natural Language
  Generation of Spanish texts},'' \emph{Expert Systems with Applications}, vol.
  120, pp. 372--386, apr 2019.

\bibitem{Jurgovsky2018}
J.~Jurgovsky, M.~Granitzer, K.~Ziegler, S.~Calabretto, P.-E. Portier,
  L.~He-Guelton, and O.~Caelen, ``{Sequence classification for credit-card
  fraud detection},'' \emph{Expert Systems with Applications}, vol. 100, pp.
  234--245, jun 2018.

\bibitem{Papernot2015}
N.~Papernot, P.~McDaniel, S.~Jha, M.~Fredrikson, Z.~B. Celik, and A.~Swami,
  ``{The Limitations of Deep Learning in Adversarial Settings},'' in
  \emph{Proceedings of the IEEE European Symposium on Security and
  Privacy}.\hskip 1em plus 0.5em minus 0.4em\relax IEEE, 2015, pp. 372--387.

\bibitem{Keshari2020}
R.~Keshari, S.~Ghosh, S.~Chhabra, M.~Vatsa, and R.~Singh, ``{Unravelling Small
  Sample Size Problems in the Deep Learning World},'' in \emph{2020 IEEE Sixth
  International Conference on Multimedia Big Data}.\hskip 1em plus 0.5em minus
  0.4em\relax IEEE, 2020, pp. 134--143.

\bibitem{Abdul-Mageed2017}
M.~Abdul-Mageed and L.~Ungar, ``{EmoNet: Fine-Grained Emotion Detection with
  Gated Recurrent Neural Networks},'' in \emph{Proceedings of the 55th Annual
  Meeting of the Association for Computational Linguistics}.\hskip 1em plus
  0.5em minus 0.4em\relax Association for Computational Linguistics, 2017, pp.
  718--728.

\bibitem{Santamaria-Granados2019}
L.~Santamaria-Granados, M.~Munoz-Organero, G.~Ramirez-Gonzalez, E.~Abdulhay,
  and N.~Arunkumar, ``{Using Deep Convolutional Neural Network for Emotion
  Detection on a Physiological Signals Dataset (AMIGOS)},'' \emph{IEEE Access},
  vol.~7, pp. 57--67, 2019.

\bibitem{Luss2014}
R.~Luss and S.~Rosset, ``{Generalized Isotonic Regression},'' \emph{Journal of
  Computational and Graphical Statistics}, vol.~23, no.~1, pp. 192--210, 2014.

\end{thebibliography}
\bibliographystyle{IEEEtran}

\EOD

\end{document}